\documentclass[reprint, amsmath, amssymb, aps, pre, floatfix, showkeys, superscriptaddress]{revtex4-1}

\usepackage{graphicx}
\usepackage{dcolumn}
\usepackage{bm}
\usepackage{color}
\usepackage{tikz}
\usepackage{pgfplots}
\usepackage{mathtools}
\usepackage{array}
\usepackage{amsmath}

\pgfplotsset{compat=1.18}

\begin{document}

\title{Identifying recurrent flows in high-dimensional dissipative chaos from low-dimensional embeddings}

\author{Pierre Beck}
\email{pierre.beck@epfl.ch}
\affiliation{
Emergent Complexity in Physical Systems Laboratory (ECPS), \'Ecole Polytechnique F\'ed\'erale de Lausanne, 1015 Lausanne, Switzerland
}

\author{Tobias M. Schneider}
\email{tobias.schneider@epfl.ch}
\affiliation{
Emergent Complexity in Physical Systems Laboratory (ECPS), \'Ecole Polytechnique F\'ed\'erale de Lausanne, 1015 Lausanne, Switzerland
}

\date{\today}

\begin{abstract}
Unstable periodic orbits (UPOs) are the non-chaotic, dynamical building blocks of spatio-temporal chaos, motivating a first-principles based theory for turbulence ever since the discovery of deterministic chaos. Despite their key role in the ergodic theory approach to fluid turbulence, identifying UPOs is challenging for two reasons: chaotic dynamics and the high-dimensionality of the spatial discretization. We address both issues at once by proposing a loop convergence algorithm for UPOs directly within a low-dimensional embedding of the chaotic attractor. The convergence algorithm circumvents time-integration, hence avoiding instabilities from exponential error amplification, and operates on a latent dynamics obtained by pulling back the physical equations using automatic differentiation through the learned embedding function. The interpretable latent dynamics is accurate in a statistical sense, and, crucially, the embedding preserves the internal structure of the attractor, which we demonstrate through an equivalence between the latent and physical UPOs of both a model PDE and the 2D Navier-Stokes equations. This allows us to exploit the collapse of high-dimensional dissipative systems onto a lower dimensional manifold, and identify UPOs in the low-dimensional embedding.
\end{abstract}

\maketitle
Driven, dissipative chaotic systems, including fluid flows, may display rich emergent phenomena, yet the mechanisms underlying those remain often unclear despite the governing equations being known. A promising approach provided by combining dynamical systems methods with advanced computational tools is based on identifying simple non-chaotic \emph{invariant solutions} embedded in the solution space of the governing nonlinear partial differential equations (PDEs). These solutions, such as equilibria or unstable periodic orbits (UPOs), are the non-chaotic building blocks supporting the chaos emerging from the deterministic PDE \cite{KawaharaAnnualRev, GrahamAnnualRev}. They capture mechanisms including self-sustaining processes of fluid turbulence \cite{Waleffe2001, Waleffe2003, Kawahara2001} and the formation of self-organized patterns \cite{Reetz2019, Hof2004}. Due to their dynamical nature, UPOs have become central to the dynamical systems approach to fluid flows \cite{KawaharaAnnualRev, Suri2020} as they recover the system's statistics  \cite{Kawahara2001, VANVEEN2006, Chandler_Kerswell_2013, PageNorgaard2024} and form a rational basis for flow control \cite{Lasagna2017}. Although UPOs have even been envisioned to yield an ergodic theory based description of turbulence \cite{Cvitanović_2013, CVITANOVIC1995}, their computation has proven challenging due to the system's chaos, and the high discretization dimension of fluid flows. 

Loop-based convergence algorithms \cite{Lan2004, Azimi2022, Parker_Schneider_2022} compute UPOs by deforming an initial time-periodic space-time field (a loop) that does not necessarily satisfy the flow equations, until its tangent vectors align with the flow at every point. Contrary to classical shooting methods \cite{Viswanath2007, Chandler_Kerswell_2013}, this circumvents time-integration, thus sidestepping exponential error amplification intrinsic to chaos, and has shown to be robust for various systems \cite{beck2024, Lasagna2017}, including the Navier-Stokes equations \cite{Ashtari_Schneider_2023, Parker_Schneider_2022}. The drawback of this approach is the high-dimensionality of the system's numerical discretization, rendering the search space for UPOs prohibitively large. 

An approach to address the dimensionality issue is to exploit the collapse of the dynamics on a low-dimensional chaotic attractor \cite{Hopf1948,Temam1990} due to dissipation. Recent advances in machine learning attempt to utilize this collapse by describing nonlinear systems in their `intrinsic' low-dimensional coordinates through nonlinear dimensionality reduction. In particular, autoencoders have proven to be efficient in nonlinear coordinate identification in fluid turbulence \cite{PageDLReview2025}. The dynamics in the autoencoder's low-dimensional latent space is often modeled with purely data-driven approaches \cite{DoanMagri2021, He_2023}, which successfully recover the statistics of the system \cite{Linot2020, Linot_Graham_2023}. However, a learned dynamics that is detached from the physical equations suggests that the internal structure of the attractor, including its UPOs, is not preserved. 
To take advantage of the low-dimensionality for computing UPOs, a reduced model must be equivalent to the original system and preserve invariant sets. If we had access to an accurate low-dimensional representation of the original system that preserves invariant sets, we could implement loop convergence algorithms directly within the low-dimensional space, solving the chaos and high-dimensionality issues at once.

\begin{figure*}
    \centering
    \includegraphics[width = 2\columnwidth]{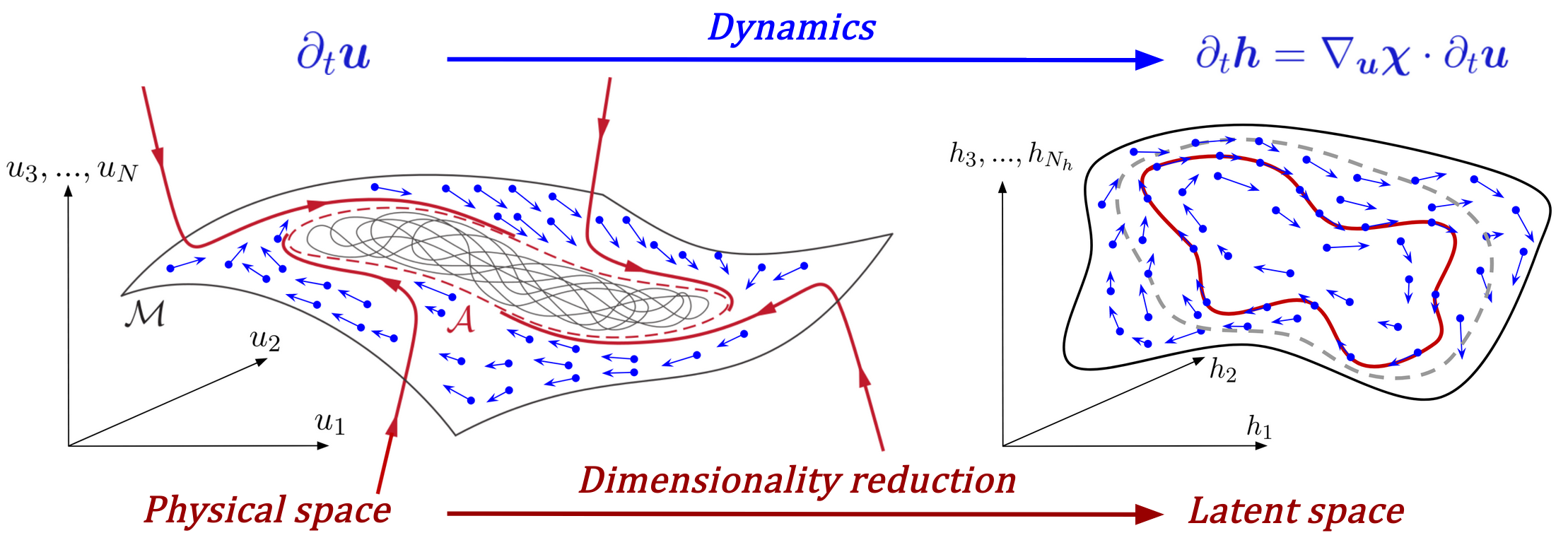}
    \caption{\textbf{Schematic representation} of the high-dimensional dynamics of a driven dissipative PDE collapsing on a chaotic attractor $\mathcal{A}$ that can be embedded in a much lower-dimensional manifold $\mathcal{M}$ (left). Through data-driven dimensionality reduction (autoencoders in our case), points from the high-dimensional physical space $\boldsymbol{u}\in\mathbb{R}^N$ are mapped to points $\boldsymbol{h}\in\mathbb{R}^{N_h}$ in the low-dimensional latent space ($N_h \ll N$). The flow vectors (blue) in the latent space are defined through chain rule, by pulling the physical flow into the latent space. Based on this flow, we implement a loop convergence algorithm, where an initial guess loop (grey dashed) is deformed into an unstable periodic orbit (red solid in the latent space). The flow vectors are always tangent to the periodic solution, which is not the case for the guess.}
    \label{fig:schematic}    
\end{figure*}

We address the challenges of UPO computations --- chaos and the high-dimensionality of spatial discretizations --- simultaneously by formulating a loop-based convergence algorithm within the latent space of an autoencoder (figure \ref{fig:schematic}). The latent dynamics required for the algorithm is derived directly from the physical equations using the chain rule \cite{Otto_Rowley_2023, LEE2020108973}, and evaluated via automatic differentiation. The inferred latent dynamics accurately reproduces statistics and is shown to preserve invariant solutions. We compute latent UPOs and confirm that they correspond to physical solutions in a one-to-one fashion. This resulting equivalence of latent and physical solutions suggest that an equivalent discretized description of the original PDE has been achieved and that the collapse of dissipative systems onto a lower-dimensional manifold can indeed be exploited in the computation of UPOs. We showcase this for a model PDE and the 2D Navier-Stokes equations.

\section*{Applications to chaotic PDEs}
An autoencoder $\boldsymbol{\mathcal{J}}$ approximates the identity, and contains a `bottleneck' layer that splits it into two parts: the encoder $\boldsymbol{\chi}$ and decoder $\check{\boldsymbol{\chi}}$, where the former maps the high-dimensional input to the low-dimensional latent space (figure \ref{fig:schematic}). We train $\boldsymbol{\mathcal{J}}$ on a physics and a tangent loss so that its outputs have accurate temporal derivatives, and its Lie derivative is equal to the flow tangent. We obtain the latent dynamics by pulling the PDE into the latent space using the chain rule. The latent loop convergence algorithm is based on this latent flow and formulated as an optimization problem. The flow vectors and the optimization updates are both evaluated using automatic differentiation. We demonstrate an equivalence between the latent UPOs and those of the original PDE for a 1D pattern-formation PDE and the 2D Navier-Stokes equations. See \textbf{Methods} for more details on $\boldsymbol{\mathcal{J}}$, data, and the convergence algorithm.

\subsection{The Kuramoto-Sivashinsky equation}
\label{sec:KS}
The Kuramoto-Sivashinsky equation (KSE) models various physical phenomena \cite{Kuramoto1976, Sivashinsky1977} and often serves as a test-bed for methods intended for the Navier-Stokes equations \cite{Azimi2022}:
\begin{equation}
    u_t + uu_x + u_{xx}  +u_{xxxx} = 0
\end{equation}
where $u$ is the velocity field on an $L$-periodic spatial domain $x\in[0,L)$, meaning $u(x,t) = u(x+L,t)$. In this formulation, $L$ is the control parameter, and for large values of $L$, the system becomes increasingly chaotic \cite{Edson2019}. The dynamics preserves anti-symmetric solutions, and we thus work in the anti-symmetric subspace $u(x) = -u(-x)$ \cite{Lasagna2017}, which discretizes the system's continuous spatial translation invariance to discrete shifts $x \to x + L/2$. The dynamics exhibits spatio-temporal chaos as shown in the top left space-time plot of panel A of figure \ref{fig:KSE_AE_recon_lat_dyn}.

\subsubsection{Reconstruction and Latent Dynamics}
\begin{figure*}
    \centering
    \includegraphics[width = 2\columnwidth]{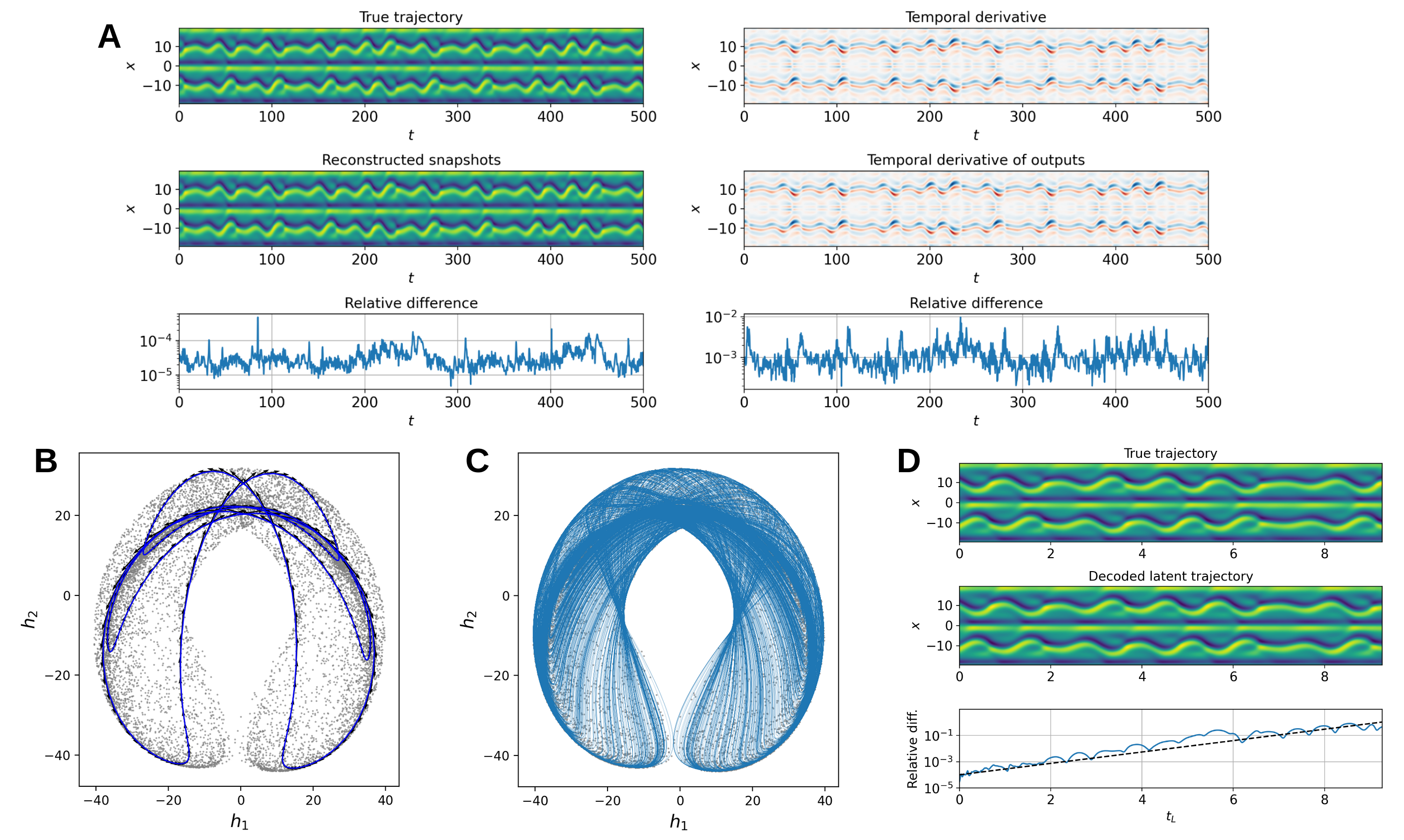}
    \caption{\textbf{Reconstruction performance of the autoencoder, and the latent dynamics} $\partial_t\boldsymbol{h}$. A typical trajectory of the KSE, simulated on a $N_x = 64$ grid, is plotted in the top left plot of panel A with the temporal derivative in the top right. The middle row shows the autoencoder output of each individual snapshot of the ground truth (left), and the temporal derivative of each output (right). Relative errors are on average $\sim\mathcal{O}(10^{-5})$ and $\sim\mathcal{O}( 10^{-3})$ respectively, and are given in log-scale in the bottom row. Panel B shows a projection on the first two latent coordinates $(h_1, h_2)$ of the latent attractor (grey) and latent flow vectors $\partial_t\boldsymbol{h}$ (black) evaluated on an encoded physical periodic orbit (blue) with period $T\approx77.4$. We observe that the flow vectors are always tangent to the periodic orbit. Integrating an arbitrary initial condition on the attractor with $\partial_t\boldsymbol{h}$ yields trajectories like the blue one in panel C. The trajectory stays on the attractor (grey background) and covers it in a stable manner. In panel D, we compare time-integration with $\partial_t\boldsymbol{u}$ (top) and $\partial_t\boldsymbol{h}$ (middle) from one same initial condition. The temporal evolution of the relative difference in the bottom (blue) matches the error growth $\propto\exp(t/t_L)$ given by the Lyapunov time $t_L \approx 21.6$ (black), thus showing no signature of inaccuracy beyond the divergence between nearby trajectories intrinsic to the chaotic system. This indicates that the latent dynamics accurately captures the original system.}
    \label{fig:KSE_AE_recon_lat_dyn}
\end{figure*}

We present results for latent dimension $N_h = 5$. The autoencoder shows solid reconstruction performance, as relative differences ${\epsilon(\boldsymbol{u},\ \check{\boldsymbol{u}})= ||\boldsymbol{u} - \check{\boldsymbol{u}}||_2 / ||\boldsymbol{u}||_2}$ for individual discretized test snapshots $\boldsymbol{u}\in\mathbb{R}^{64}$ are on average $3\times 10^{-5}$. The inclusion of physical and tangent loss terms in the autoencoder's training drive the relative errors $\epsilon(\partial_t\boldsymbol{u},\ \partial_t\check{\boldsymbol{u}})$ between the temporal derivatives of the test snapshots and of the autoencoder output to also be small (panel A of figure \ref{fig:KSE_AE_recon_lat_dyn}), with an average of $1\times 10^{-3}$, demonstrating weak spectral bias \cite{Xu_spectralbias_2020}. First order derivatives are thus mapped accurately, meaning the latent dynamics $\partial_t\boldsymbol{h} = \nabla_{\boldsymbol{u}}\boldsymbol{\chi}\cdot\partial_t\boldsymbol{u}$ is tangent to encoded physical trajectories in the latent space --- a necessary condition for a coherent latent dynamics. For example, the latent flow vectors computed through automatic differentiation at points of an encoded physical UPO are always tangent to it (panel B of figure \ref{fig:KSE_AE_recon_lat_dyn}). Importantly, the $\partial_t\boldsymbol{h}$ vectors were evaluated using the decoded states $\check{\boldsymbol{u}}$ of the UPO rather than the original $\boldsymbol{u}$.

Time-integrating the latent dynamics $\partial_t\boldsymbol{h}$ from an arbitrary initial condition is stable and indicates ergodicity, as the trajectory gradually covers the latent attractor and stays on it for all times (panel C of figure \ref{fig:KSE_AE_recon_lat_dyn}). We compare two trajectories obtained by separately integrating the same initial condition using $\partial_t\boldsymbol{u}$ and $\partial_t\boldsymbol{h}$, with the time-evolution of the relative error between the trajectories serving as an accuracy test. We compare this deviation to the accuracy limit prescribed by the exponential error growth of the system, where the Lyapunov time $t_L \approx 21.6$ is the benchmark. The deviation grows at the same rate, thus showing no signs of inaccuracy beyond the divergence between nearby trajectories intrinsic to the chaotic system (panel D of figure \ref{fig:KSE_AE_recon_lat_dyn}).

\subsubsection{Unstable Periodic Orbits}

\begin{figure*}
    \centering
    \includegraphics[width = 2\columnwidth]{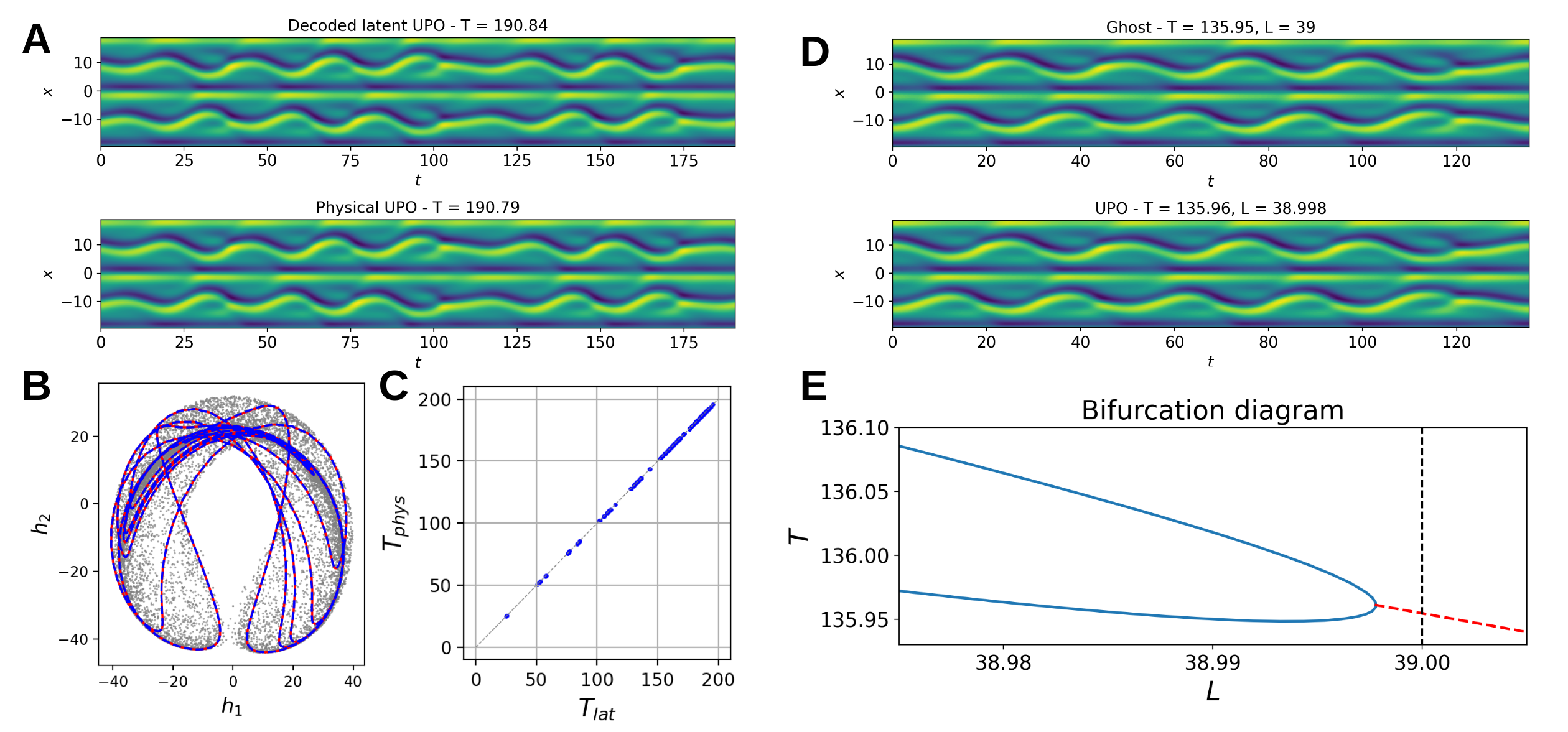}
    \caption{\textbf{Equivalence between latent UPOs and physical UPOs in the KSE}. Panel A shows an example of a decoded latent UPO (top) with period $T_{lat} \approx 190.84$ and the physical UPO it converges to (bottom) with period $T_{phys} \approx 190.79$. A latent projection of the UPO pair is shown in panel B, with the latent chaotic attractor in grey, the latent UPO in red and the encoded physical UPO in blue dashed. There are no visible differences in both the physical and latent representations. The periods of all unique 195 UPO pairs agree very well, as shown by the $y = x$ trend in panel C, with an average relative difference between $T_{phys}$ and $T_{lat}$ of $\approx2.4\times10^{-4}$. In 16 cases, the latent UPO does not converge to a physical UPO, but to a very low local minimum of the cost function. Slight perturbation of the control parameter $L$ results in a UPO as shown in panel D. Through pseudo-arclength continuation we obtain the bifurcation diagram of panel E, which shows the saddle-node bifurcation (blue) and the resulting ghost state (the local minimum of the cost function --- red dashed) passing through the original parameter $L=39$ (black dashed).}
    \label{fig:KSE_upos}
\end{figure*}

We identify latent UPOs by implementing a loop convergence algorithm directly within the latent space based on  $\partial_t\boldsymbol{h}$. The latent cost function ${J_{upo}^2 = \int_0^1\rvert\rvert \partial_t\boldsymbol{h} - T^{-1}\partial_s\boldsymbol{h}\lvert\lvert^2ds}$ quantifies the misalignment between the flow and tangent vectors at every point of a loop $\boldsymbol{h}(s)$, where $ s\in[0, 1)$ parametrizes the loop. Zeros of this non-negative cost function correspond to latent UPOs (or physical UPOs for the physical analogue of the cost function). We find these roots through an optimization process, for which gradient information is obtained using automatic differentiation. The optimization requires an initial guess, which we obtain by monitoring close recurrences in the latent dynamics, marking possible shadowings of UPOs. From one latently integrated trajectory with $t_{max} = 10^5$, such a latent recurrent flow analysis (RFA) yields 1289 guesses. These guesses converge to 1157 latent UPOs, 211 of which are unique. A necessary condition for the latent dynamics to be an accurate representation of the physical dynamics is for the latent UPOs to correspond to physical ones. To verify this, we decode the latent UPOs and use them as initial guesses in a physical convergence algorithm. Indeed, from the 211 latent solutions, we obtain 195 unique physical UPOs (or 167 after accounting for the system's discrete shift symmetry $x\mapsto x + L/2$). The equivalence between latent and physical solutions is highlighted by the relationship between their periods. The relative difference $\epsilon(T_{phys},\ T_{lat})$ is on average $2\times10^{-4}$, with no observed growth for longer UPOs (panel C of figure \ref{fig:KSE_upos}). For example, a long latent UPO of period $T_{lat}\approx 190.84$ converges to a physical UPO with $T_{phys}\approx 190.79$. Their identical structure can be seen in the physical space-time plots, and the latent projection of figure \ref{fig:KSE_upos}. These observations indicate that the simple invariant sets are preserved in the latent space, with an equivalent low-dimensional dynamics.

The 16 latent UPOs that do not converge to physical solutions correspond to local minima of the physical cost function, with values very close to 0. Slight perturbations to the control parameter $L$ result in all 16 converging. In each case, we identify a saddle-node bifurcation with critical bifurcation parameter $L^*$ very close to $L=39$. Thus, $L^*$ slightly changed for the latent space as often observed in changes in discretization. The local minima can be interpreted as `ghosts states' of saddle-node bifurcations \cite{ghosts}. The strong similarity between a ghost state and the UPO at the saddle-node bifurcation, and a bifurcation diagram highlighting how close the saddle-node bifurcation happens to $L=39$, are shown in panels D and E of figure \ref{fig:KSE_upos}.
Thus invariant sets of the KSE are preserved by the latent dynamics despite a reduction of the directization dimension by an order of magnitude and UPOs can reliably be computed in the only $N_h = 5$-dimensional latent space.

\subsection{Kolmogorov Flow}
\label{sec:Kolmogorov}
We now conduct the same analysis for a fluid flow that is well-studied, both numerically \cite{Chandler_Kerswell_2013, LucasYasuda2022, Parker_Schneider_2022} and experimentally \cite{Suri2020}. We consider the 2D incompressible Navier-Stokes equations in vorticity formulation
\begin{equation}\label{eqn:vorticity_eqn}
    \partial_t \omega + u\partial_x \omega + v\partial_y \omega = Re^{-1} \nabla ^ 2 \omega + 4\cos 4 y
\end{equation}
where $\omega = (\boldsymbol{\nabla}\times\boldsymbol{u})\cdot\hat{\boldsymbol{z}}$ is the vorticity field, $Re$ is the Reynolds number, and the domain is $(x,y)\in[0,2\pi)^2$ with periodic boundary conditions. The system has discrete symmetries, namely rotation by $\pi$, and the shift-reflect symmetry $(x,y) \mapsto (-x, y + \pi / 4)$. Moreover, the system possesses continuous translation symmetry $x \mapsto x + s,\; s\in\mathbb{R}$. Along with the extra spatial dimension and the non-locality of the vorticity equation, this results in an additional technical difficulty compared to the KSE, since an infinity of snapshots are dynamically equivalent up to an arbitrary phase-shift $\phi$ in the $x$-direction. We apply First Fourier Slicing \cite{Rowley2000, Marensi2023, Engel2023} to factor out $\phi$ and work with vorticity fields with a fixed phase $\phi = 0$ before training the autoencoder (see \textbf{Methods}). This $\phi$ is fully interpretable and its evolution $\partial_t\phi$ can be recovered from the `sliced' snapshots using the reconstruction equations \cite{Rowley2000}. For what follows, we consider temporal derivatives $\partial_t\omega$ to be `within the slice' where $\phi$ remains factored out. 

We set $Re = 40$, at which the system exhibits turbulent behavior \cite{Chandler_Kerswell_2013}. We simulate the system over a $(N_x, N_y) = (64, 64)$ grid, using {\fontfamily{qcr}\selectfont jax-cfd}'s spectral solver \cite{Kochkov2021-ML-CFD, Dresdner2022-Spectral-ML}, resulting in real vorticity fields $\boldsymbol{\omega}\in\mathbb{R}^{64\times 64}$. This resolution is large enough to give accurate results, without over-resolving the system \cite{Parker_Schneider_2022}. 

\subsubsection{Reconstruction and Latent Dynamics}
With the chosen resolution, the input vector reduces to 1847 variables after accounting for the fixed $\phi$, symmetries in the Fourier coefficients from $\boldsymbol{\omega}$ being purely real, and some Fourier coefficients being zero from de-aliasing. Autoencoders with $N_h \in \{64, 128\}$ resulted in inaccurate temporal derivatives of the outputs $\partial_t\check{\boldsymbol{\omega}}$. Networks with $N_h \in \{ 192,256\}$ showed satisfactory performance, causing us to work with $N_h = 192$, an order of magnitude smaller than the spatial discretization dimension. This $N_h$ also agrees with recent estimates on attractor dimension for this system \cite{cleary2025}. We do not claim that $N_h = 192$ is minimal, rather we chose it for efficient FFTs.

\begin{figure*}
    \centering
    \includegraphics[width = 2\columnwidth]{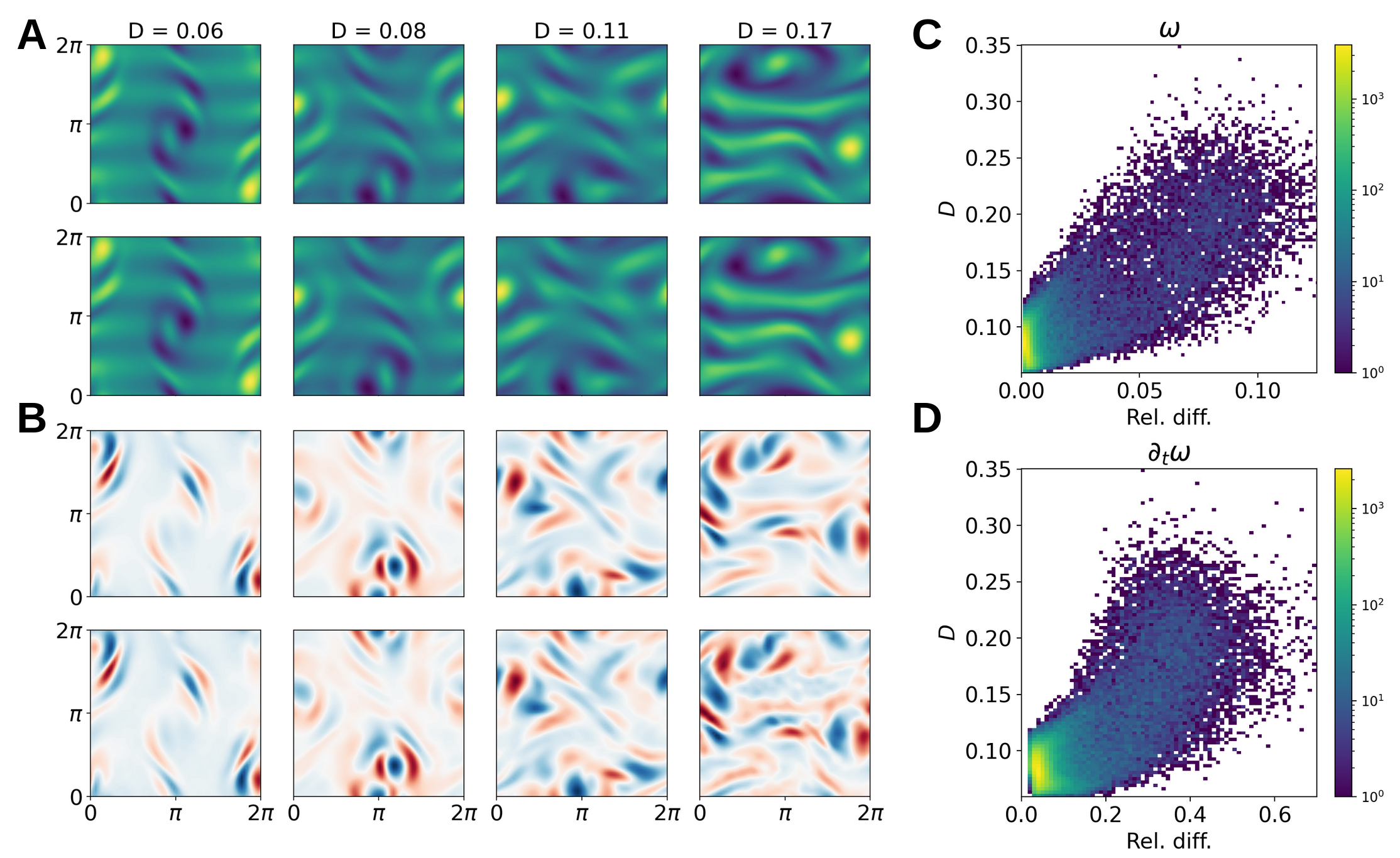}
    \caption{\textbf{Reconstruction performance of the autoencoder for Kolmogorov flow}. Panel A shows test snapshots (top) and the autoencoder output (bottom), ordered in increasing dissipation $D$ (left to right). The relative differences $\epsilon(\boldsymbol{\omega}, \check{\boldsymbol{\omega}})$ are on average 1\%, and 0.16\%, 0.14\%, 0.71\% and 4.4\% for the individual snapshots. The temporal derivative for the same snapshots and the autoencoder outputs are shown in panel B. Visual differences appear only for the high dissipation snapshot. The relative differences $\epsilon(\partial_t\boldsymbol{\omega}, \partial_t\check{\boldsymbol{\omega}})$ are 8\% on average, and 2.98\%, 4.10\%, 7.04\% and 20.8\% for the individual snapshots. Histograms \cite{cleary2025} of the normalized dissipation against $\epsilon(\boldsymbol{\omega}, \check{\boldsymbol{\omega}})$ and $\epsilon(\partial_t\boldsymbol{\omega}, \partial_t\check{\boldsymbol{\omega}})$ are displayed in panels C and D.}
    \label{fig:Kol_reconstruction}
\end{figure*}

The relative reconstruction error $\epsilon(\boldsymbol{\omega},\ \check{\boldsymbol{\omega}})$ of the autoencoder for test vorticity snapshots is on average 1\%, indicating good reconstruction performance. This is illustrated for snapshots of varying normalized instantaneous dissipation rates $D$ in panel A of figure \ref{fig:Kol_reconstruction}. The mean relative error of temporal derivatives $\epsilon(\partial_t\boldsymbol{\omega},\ \partial_t\check{\boldsymbol{\omega}})$ over the test set is 8\%. The temporal derivatives $\partial_t\boldsymbol{\omega}$ and $\partial_t\check{\boldsymbol{\omega}}$ of the same snapshots as before are plotted in panel B. Only for the high-dissipation snapshots do visual differences appear, while still maintaining the general shape. This is also confirmed in the 2D histograms of figure \ref{fig:Kol_reconstruction}, which show $\epsilon(\boldsymbol{\omega},\ \check{\boldsymbol{\omega}})$ and $\epsilon(\partial_t\boldsymbol{\omega},\ \partial_t\check{\boldsymbol{\omega}})$ against $D$: the performance is generally good, but decreases for the rarer high-dissipation events. 

\begin{figure*}
    \centering
    \includegraphics[width = 2\columnwidth]{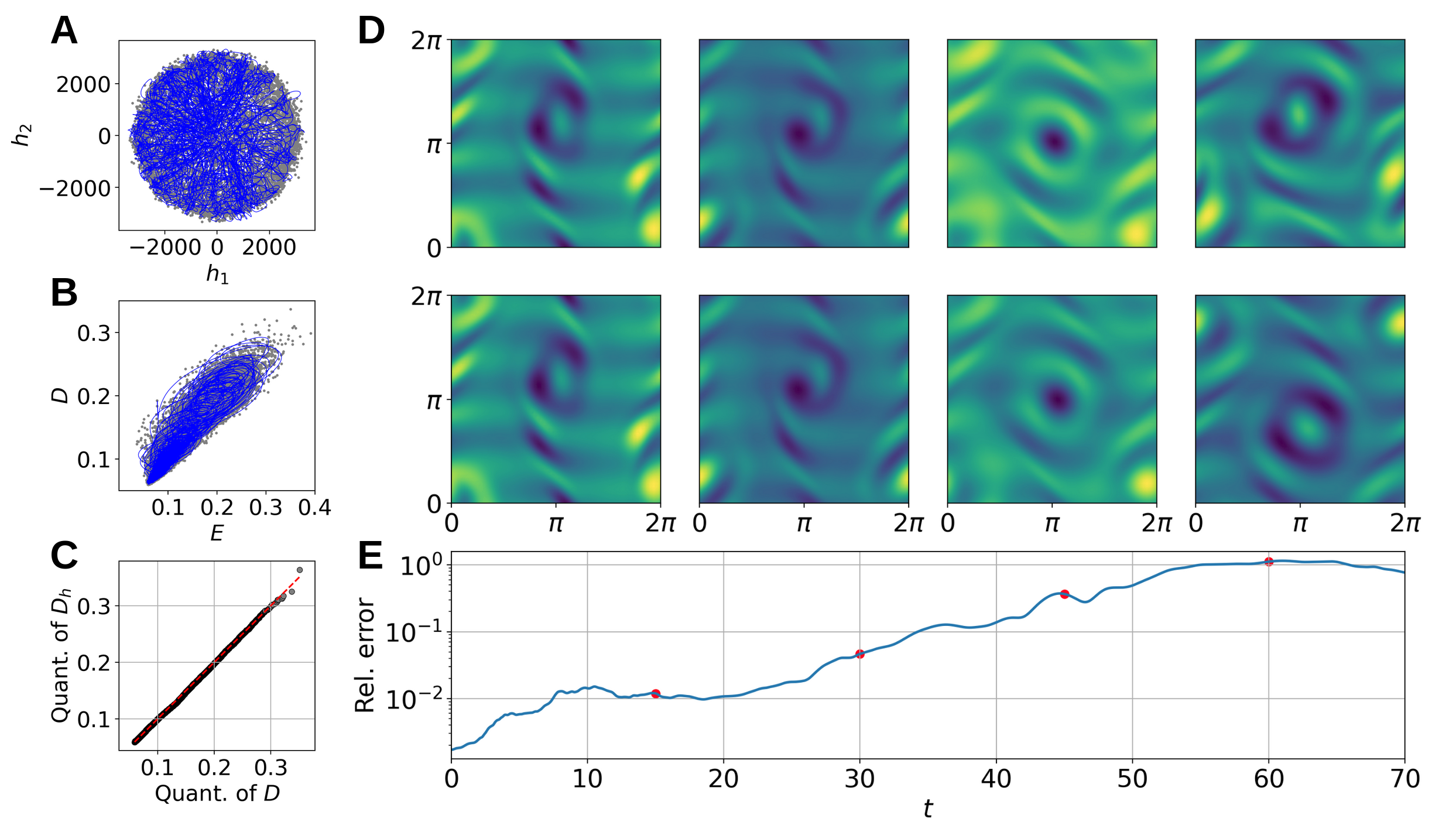}
    \caption{\textbf{Latent dynamics} $\partial_t\boldsymbol{h}$ \textbf{of the autoencoder}. Encoded physical snapshots (grey) and a trajectory integrated with $\partial_t\boldsymbol{h}$ (blue) in the latent space coordinates $(h_1, h_2)$ are depicted in panel A. The latent attractor shows an octogonal shape in this projection, relating to the discrete symmetries of the system \cite{PBK2021}. The latent trajectory does not leave the attractor and starts covering it, indicating ergodicity. A similar conclusion can be drawn from the physical projection onto normalized dissipation $D$ and energy input $E$, shown in panel B. In particular, the dissipations of physical snapshots and decoded latent trajectories, $D$ and $D_h$ respectively, agree in distribution, indicated by the $y=x$ trend of the quantile-quantile plot of panel C. Snapshots of trajectories integrated with $\partial_t\boldsymbol{\omega}$ and $\partial_t\boldsymbol{h}$ from one same initial condition for $t \in\{15,30,45,60\}$ are shown in panel D, with the evolution of the relative error $\epsilon(\boldsymbol{\omega},\ \check{\boldsymbol{\omega}})$ shown in panel E. The red points mark the errors for the plotted snapshots.}
    \label{fig:Kol_lat_dyn}
\end{figure*}

We time-integrate the latent dynamics $\partial_t\boldsymbol{h}$ over 4000 time units to obtain the trajectory shown in panel A of figure \ref{fig:Kol_lat_dyn}. Judging by the $(h_1,h_2)$-projection, we observe ergodic dynamics without numerical instabilities as the trajectory stably covers the attractor without leaving it. This indicates that the attractor is indeed an invariant set, despite the autoencoder's slightly larger reconstruction errors compared to the KSE. We verify that the statistics of integrated physical quantities agree by decoding the trajectory to the physical space, and projecting it into the $(E,D)$ plane, where $E$ is the normalized energy input rate. We observe a similar covering, and, in fact, the distributions of dissipations computed from physical snapshots $D$, and decoded latent snapshots $D_h$, are almost exactly the same as highlighted by the $y = x$ trend in the quantile-quantile plot of figure \ref{fig:Kol_lat_dyn}. Panel D of figure \ref{fig:Kol_lat_dyn} compares trajectories integrated from one same initial condition with the physical dynamics $\partial_t\boldsymbol{\omega}$ and latent dynamics $\partial_t\boldsymbol{h}$. The latently integrated trajectory follows the physical one for a while before diverging as highlighted by their relative difference in panel E.

\subsubsection{Unstable Periodic Orbits}

\begin{figure*}
    \centering
    \includegraphics[width = 2\columnwidth]{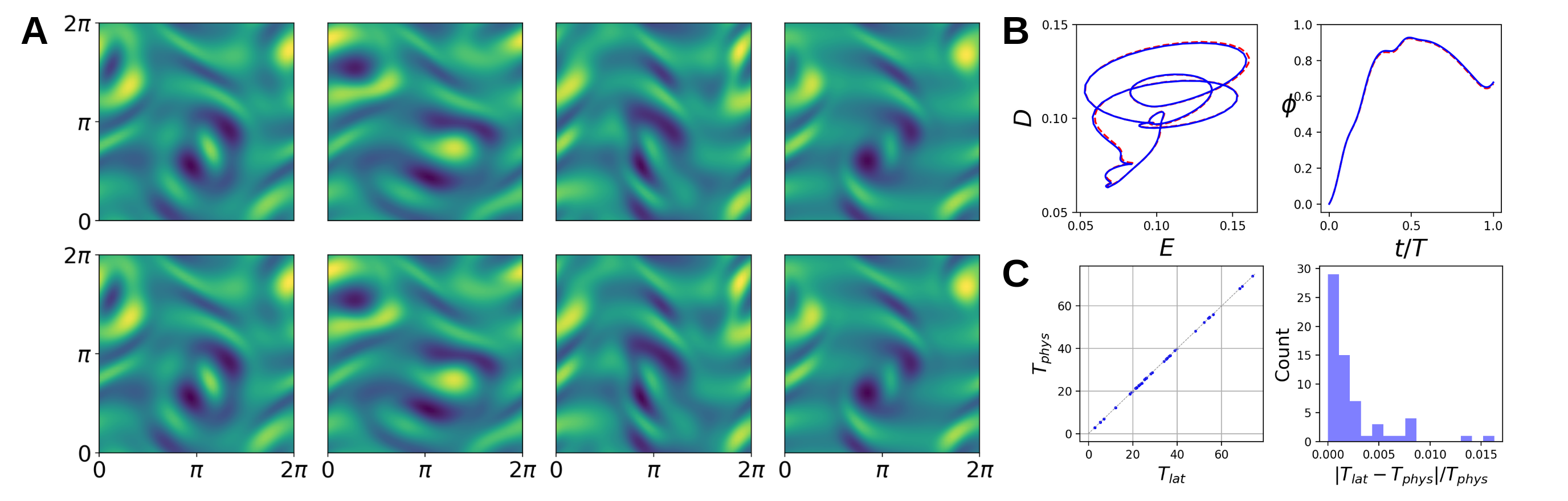}
    \caption{\textbf{Equivalence of latent and physical invariant solutions}. Panel A shows a pair of a decoded latent UPO (top) with $T_{lat} = 22.60$, and the physical UPO it converges to with $T_{phys} = 22.62$, plotted at times $n T/4$, where $n = 0,1,2,3$ (columns 1 to 4). Animations of this UPO pair, and two additional ones, are available in the \textbf{Supplementary Information}. Projections in the $(E,D)$ plane of the decoded latent UPO (red dashed) and the physical UPO (blue) are shown on the left of panel B, while the evolution of the phase $\phi$ (same color coding), obtained from the reconstruction equations, is shown on the right. Panel C confirms that the periods of the latent UPOs and those of the physical UPOs they converged to agree very well (left), with $\epsilon(T_{phys}, \; T_{lat})$ on average $2\times10^{-3}$ (right).}
    \label{fig:Kol_invariant_sol}
\end{figure*}

We proceed similarly as for the KSE to compute latent UPOs. After formulating the loop convergence algorithm based on $\partial_t\boldsymbol{h}$, we generate guesses for the algorithm from 38 different RFAs. Twenty-nine were focused on shorter recurrences, while 9 were targeted at longer guesses. The short searches yield 1016 guesses, 184 of which converged to latent UPOs. The long searches generate 129 guesses, with 15 convergences. We obtain 69 unique latent UPOs. To check that the latent UPOs correspond to physical ones, we decode and re-converge them in the physical space. This results in 63 physical UPOs, or 39 unique groups of UPOs when accounting for discrete symmetries. For example, we find all 8 symmetric versions of the well-known physical UPO with period $T_{phys} \approx 5.38$ (`P1' in \cite{Chandler_Kerswell_2013}). One such decoded latent UPO and its physical counterpart is shown in panel A of figure \ref{fig:Kol_invariant_sol}, with snapshots plotted at quarter period intervals. The pair has periods $(T_{lat},\ T_{phys}) = (22.60,\ 22.62)$. There is a clear similarity between the latent and physical UPOs, which can also be seen in the $(E,D)$ plane, or from the evolution of the phase $\phi$ obtained via the reconstruction equations in panel B. We find good agreement between the periods of the latent UPOs and their physical counterparts, with $\epsilon(T_{phys},\ T_{lat})$ on average $2.4\times10^{-3}$. This is highlighted in panel C of figure \ref{fig:Kol_invariant_sol}: plotting $T_{phys}$ against $T_{lat}$ for the 63 pairs shows a clear $y = x$ trend, and a histogram of the distribution of $\epsilon(T_{lat},\ T_{phys})$ confirms that it is generally very small. We find UPOs of both short and long periods, ranging from 2.92 up to 74.07. To the best of our knowledge, this is the longest UPO found for this system, at least when comparing to \cite{Chandler_Kerswell_2013, PageNorgaard2024, Parker_Schneider_2022, Redfern2024}, highlighting the power and robustness of a variational loop convergence algorithm formulated directly in a low-dimensional latent space. In 6 cases the decoded latent UPO does not converge to a physical UPO. However, for all 6, perturbations in $Re$ lead to converged UPOs and we are able to identify the relevant saddle-node bifurcation, confirming that these are ghost states \cite{ghosts}.

\section*{Discussion, Impact and Outlook}
While non-chaotic unstable periodic orbits (UPOs) support chaos in driven-dissipative systems and carry the promise to unlock a path towards a still elusive first-principle description of fluid turbulence, they remain challenging to identify computationally due to the defining property of chaos, exponential error amplification, and the high-dimensionality of the spatial discretization required. 
We address both challenges simultaneously by formulating a loop-based convergence algorithm within a low-dimensional embedding of the chaotic attractor for the Kuramoto-Sivashinsky model and the 2D Navier-Stokes equations. The embeddings are obtained in a data-driven fashion using a basic autoencoder neural network, trained on Fourier coefficients of the physical fields, with a loss function appropriately designed to encode the equations of motion and to preserve the flow's tangent spaces, as necessary for a first-order-in-time dynamical system. The latent dynamics required for the convergence algorithm is derived directly from the physical equations by chain rule and evaluated using automatic differentiation. For a perfect embedding, this dynamics is equivalent to the physical system by construction. Despite the autoencoder giving only an approximate embedding, we find that the learned mapping also preserves the original system. In particular, distributions of physical quantities computed from the integrated latent dynamics match those of the original physical dynamics.

Beyond statistics, the mappings also preserve the internal dynamical structure of the attractor. Specifically, UPOs, which form building blocks of the dynamics and define the most relevant simple invariant sets, are mapped one-to-one. We verified this correspondence through the loop convergence algorithm, which identifies latent UPOs through an optimization procedure that circumvents time-marching the chaotic dynamics with updates obtained via automatic differentiation. The resulting latent solutions are then confirmed to correspond to physical ones by decoding and re-converging the latent UPOs in physical space. Together with the statistics-preserving latent ergodic dynamics, this evidences the preservation of invariant sets --- both simple ones and the entire attractor --- by the learned embedding. 

As the learned embedding preserves UPOs, the search for UPOs can now be entirely formulated in the low-dimensional latent space, from generating guesses to converging them. While the data-driven dimensionality reduction may also be used in conjunction with a classical shooting algorithm, it is especially beneficial for the variational loop convergence algorithm, as the optimization operates over entire space-time fields, increasing the number of degrees of freedom immensely. Employing the algorithm in the low-dimensional latent space decouples the attractor dimension from the discretization dimension and significantly reduces the search space for UPOs. This allows for the computation of UPOs otherwise inaccessible in high-dimensional nonlinear systems, such as the long UPOs of Kolmogorov flow computed here.

We have shown how appropriately designed data-driven dimensionality reduction in combination with variational convergence algorithms can break the curse of dimensionality and simultaneously avoid instabilities due to exponential error amplification, thereby allowing us to efficiently identify non-chaotic simple invariant solutions underpinning chaos. 
While our focus is mainly conceptual --- the appropriate autoencoder loss, deriving the latent dynamics, and using this to implement a latent convergence algorithm --- we see much untapped potential in the optimization of the autoencoder architecture. This is especially important when extending these methods to  3D fluid flows, for example by directly capturing the system's discrete symmetries in the architecture.
In general, this work showcases how the collapse on a low-dimensional manifold related to the dissipative nature of many high-dimensional problems can be exploited in practice. The combination of modern machine-learning tools and differentiable code with classical dynamical systems methods for invariant solutions promises to advance the ergodic theory approach to fluid turbulence and may eventually open avenues towards the first-principles based theory to turbulent flows that has been envisioned since the identification of deterministic chaos in the mid-20th century.

\section*{Methods}
\label{sec:methods}

\subsection*{Latent Dynamics}
Consider an ODE $\partial_t \boldsymbol{u} = \boldsymbol{F} (\boldsymbol{u})$ for $\boldsymbol{u}\in \mathbb{R}^N$ and suppose we have mappings ${\boldsymbol{\chi}:\mathbb{R}^N\rightarrow\mathbb{R}^{N_h}}$ and $\boldsymbol{\check{\chi}}:\mathbb{R}^{N_h}\rightarrow\mathbb{R}^N$, with $N_h < N$ and ${\boldsymbol{\check\chi}\circ\boldsymbol{\chi}(\boldsymbol{u}) = \boldsymbol{u}}$. We refer to $\mathbb{R}^N$ as the \textit{physical space}, and $\mathbb{R}^{N_h}$ as the \textit{latent space}. The chain rule yields the latent dynamics
\begin{equation}
\label{eqn:latent_dyn}
    \partial_t \boldsymbol{h} = \nabla_{\boldsymbol{u}} \boldsymbol{\chi}\cdot\partial_t \boldsymbol{u}= \nabla_{\boldsymbol{u}} \boldsymbol{\chi}\cdot \boldsymbol{F} (\boldsymbol{u})
\end{equation}
where $\boldsymbol{h}\in\mathbb{R}^{N_h}$. In general we only have access to ${\check{\boldsymbol{u}} = \check{\boldsymbol{\chi}}(\boldsymbol{h})}$, so equation \ref{eqn:latent_dyn} should be written as
\begin{equation}
\label{eqn:latent_dyn_actual}
    \partial_t \boldsymbol{h} = \nabla_{\boldsymbol{u}} \boldsymbol{\chi}\big\rvert_{\boldsymbol{u} = \check{\boldsymbol{u}}}\cdot \boldsymbol{F} (\check{\boldsymbol{u}})
\end{equation}
This detail is redundant for perfect mappings $\boldsymbol{\chi}$ and $\check{\boldsymbol{\chi}}$. Usually however, we cannot obtain perfect mappings and rely on approximations. 

\subsection*{Autoencoder architecture}
To reduce the high-dimensional physical space, we use an autoencoder, which is a neural network that consists of two parts, namely the \textit{encoder} $\boldsymbol{\chi}: \mathbb{R}^N\rightarrow\mathbb{R}^{N_h}$ and the \textit{decoder} $\check{\boldsymbol{\chi}}: \mathbb{R}^{N_h}\rightarrow\mathbb{R}^N$. Given input data $\boldsymbol{u}\in \mathbb{R}^N$, we train the autoencoder $\boldsymbol{\mathcal{J}} = \check{\boldsymbol{\chi}}\circ\boldsymbol{\chi}$ so that approximately ${\boldsymbol{\mathcal{J}}(\boldsymbol{u}) \approx \boldsymbol{u}}$, with $N_h \ll N$. The architecture we use is that of a `hybrid neural network' \cite{Linot2020, Linot_Graham_2023}, where the dimensionality reduction is equivalent to projection onto the first $N_h$ POD modes at linear order, and neural networks model a non-linear correction:
\begin{align*}
    \boldsymbol{h} &= \boldsymbol{\chi}(\boldsymbol{u}) = \boldsymbol{U}_{N_h}^T\boldsymbol{u} + \boldsymbol{\mathcal{E}}(\boldsymbol{U}^T\boldsymbol{u}) \\
    \boldsymbol{U}^T\check{\boldsymbol{u}} &= \check{\boldsymbol{\chi}}(\boldsymbol{h}) = (\boldsymbol{h}, \boldsymbol{0})^T + \boldsymbol{\mathcal{D}}(\boldsymbol{h})
\end{align*}
The columns of the matrix $\boldsymbol{U}$ correspond to the POD modes, while $\boldsymbol{U}_{N_h}$ only contains the first $N_h$ modes. The vector functions $\boldsymbol{\mathcal{E}}$ and $\boldsymbol{\mathcal{D}}$ are standard dense neural networks whose parameters are optimized during the training process. We implement $\boldsymbol{\mathcal{E}}$ and $\boldsymbol{\mathcal{D}}$ with {\fontfamily{qcr}\selectfont Flax} \cite{flax2020github}, a neural network library for {\fontfamily{qcr}\selectfont JAX} \cite{jax2018github}, and optimize them using {\fontfamily{qcr}\selectfont Optax} \cite{deepmind2020jax}. The latent dynamics is thus 
\begin{equation} \label{eqn:expl_dyn}
\partial_t\boldsymbol{h} = \Big( \boldsymbol{U}^T_{N_h} + \textbf{J}_{\boldsymbol{\mathcal{E}}}(\check{\boldsymbol{u}})\cdot\boldsymbol{U}^T\Big)\cdot\partial_t\check{\boldsymbol{u}}
\end{equation}
where $\textbf{J}_{\boldsymbol{\mathcal{E}}} = \nabla_{\boldsymbol{u}} \boldsymbol{\mathcal{E}}$ is the Jacobian of $\boldsymbol{\mathcal{E}}$. We time-integrate the latent dynamics using {\fontfamily{qcr}\selectfont JAX}'s {\fontfamily{qcr}\selectfont odeint} function, which is an adaptive stepsize (Dormand-Prince) Runge-Kutta scheme \cite{DormandPrince}.

\subsection*{Loss functions}
\textbf{Tangent loss}: The autoencoder approximates the identity for points on the attractor. Therefore, for attractor points perturbed in directions that stay on the attractor, the autoencoder should still serve as the identity. One direction that always stays on the attractor is the flow direction, namely the unit tangent $\hat{\boldsymbol{t}} = \partial_t \boldsymbol{u}/|| \partial_t \boldsymbol{u} ||_2$. The evolution for time $t$ in the $\hat{\boldsymbol{t}}$ direction is given by the flow function $\boldsymbol{\Phi}^t$ and stays on the attractor. For an accurate autoencoder, we then require
\begin{equation}
\boldsymbol{\mathcal{J}}(\boldsymbol{\Phi}^t(\boldsymbol{u})) = \boldsymbol{\Phi}^t(\boldsymbol{u})
\end{equation}
The Lie derivative (or directional derivative) of the autoencoder is given by
\begin{align}
    \nabla_{\dot{\boldsymbol{u}}} \boldsymbol{\mathcal{J}}(\boldsymbol{u}) &= \lim_{t' \rightarrow 0}\frac{1}{t'} [\boldsymbol{\mathcal{J}}(\boldsymbol{u}(t + t')) - \boldsymbol{\mathcal{J}}(\boldsymbol{u}(t))] \\
    &= \lim_{t' \rightarrow 0}\frac{1}{t'} [\boldsymbol{\mathcal{J}}(\boldsymbol{\Phi}^{t'}(\boldsymbol{u}(t))) - \boldsymbol{\mathcal{J}}(\boldsymbol{u}(t))]\\
    &=\lim_{t' \rightarrow 0}\frac{1}{t'} [\boldsymbol{\Phi}^{t'}(\boldsymbol{u}(t)) - \boldsymbol{u}(t)] \\
    \implies \nabla_{\dot{\boldsymbol{u}}} \boldsymbol{\mathcal{J}}(\boldsymbol{u}) &=\partial_t\boldsymbol{u}
\end{align}
It follows that
\begin{equation}
    \hat{\boldsymbol{t}} \cdot \nabla \boldsymbol{\mathcal{J}} = \hat{\boldsymbol{t}} 
\end{equation}
Hence the tangent vector is an eigenvector with eigenvalue 1 of the transpose of the Jacobian of the autoencoder. This allows us to formulate the tangent loss:
\begin{equation}
    \mathcal{L}_{tang} = ||\hat{\boldsymbol{t}} \cdot \nabla \boldsymbol{\mathcal{J}} -  \hat{\boldsymbol{t}} ||_2
\end{equation}
We practically compute this loss with a finite difference approximation. \citet{Fainstein2025} investigate a similar loss in the Lorenz system. \\
\newline
\textbf{Other losses and total loss}: We include a physics loss to reduce the spectral bias:
\begin{equation}
    \mathcal{L}_{phys} = \rvert\rvert\partial_t \boldsymbol{u} - \partial_t \check{\boldsymbol{u}} \lvert\lvert_2 / \rvert\rvert\partial_t \boldsymbol{u}\lvert\lvert_2
\end{equation}
Note that including a physics loss does not always resolve spectral bias, and other methods . In our case, we found that learning the Fourier coefficients instead of the spatial discretization variables, together with $\mathcal{L}_{phys}$ and $\mathcal{L}_{tang}$ significantly reduced spectral bias.

Splitting $\boldsymbol{\chi}$ into linear (POD) and nonlinear (neural network) parts gives rise to a necessary condition of recovering the first $N_h$ POD modes in the output \cite{Linot2020, Linot_Graham_2023}, defining a POD loss:
\begin{equation}
    \mathcal{L}_{POD} = ||\boldsymbol{\mathcal{D}}_{N_h}(\boldsymbol{h}) + \boldsymbol{\mathcal{E}}(\boldsymbol{U}^T\boldsymbol{u)}||_2
\end{equation}
where $\boldsymbol{\mathcal{D}}_{N_h}$ is the first ${N_h}$ components of $\boldsymbol{\mathcal{D}}$. This may be interpreted as the neural network not interfering with the linear decomposition. Finally, we also use a standard relative loss between the autoencoder input and output, as well as their squares:  
\begin{equation}
    \mathcal{L}_{rel} =||\boldsymbol{u} - \check{\boldsymbol{u}}||_2/||\boldsymbol{u}||_2 \quad \mathrm{and} \quad \mathcal{L}_{sq} = ||\boldsymbol{u}^2 - \check{\boldsymbol{u}}^2||_2/||\boldsymbol{u}^2\lvert\lvert_2
\end{equation}
Including the loss of the squares helps with learning the rare bursting events of chaotic fluid flows, instead of re-weighting the data \cite{PageNorgaard2024}. The final loss function is a linear combination of the above losses:
\begin{equation}
    \mathcal{L} = \mathcal{L}_{rel} + \mathcal{L}_{sq} + \lambda_1\mathcal{L}_{POD} + \lambda_2\mathcal{L}_{tang} + \lambda_3 \mathcal{L}_{phys} + \lambda_4 \mathcal{L}_{reg}
\end{equation}
where $\lambda_i$ are scalar coefficients and $\mathcal{L}_{reg}$ is a regularization term to prevent over-fitting. Note that in theory it should be enough to only use an input-output loss $\mathcal{L}_{rel}$ to train our networks. Including the other loss terms accelerates the learning process towards desirable parameters.

\subsection*{Data and Training}
\textbf{Kuramoto-Sivashinsky}:
We generate data on a spatial grid with $N_x = 64$ points using the ETDRK4 scheme \cite{Kassam2005}. The autoencoder is trained on the Fourier components of $\boldsymbol{u}\in\mathbb{R}^{64}$, namely the spatial FFT $\tilde{\boldsymbol{u}} = \mathcal{F}_x(\boldsymbol{u})$. Since $\boldsymbol{u}$ is real, we discard the negative frequencies of $\tilde{\boldsymbol{u}}$, as they are complex conjugates of the positive ones. Moreover, since we work in the anti-symmetric subspace, the Fourier modes are purely imaginary $\tilde{\boldsymbol{u}} = i\boldsymbol{v}$. With standard 2/3 de-aliasing, this results in 21 unique degrees of freedom, defining the input vectors to the encoder $\boldsymbol{\chi}$ to be the first 21 non-zero components of $\boldsymbol{v}$. 

In the training process, we set $\lambda_1 = \lambda_2 = 1$, while $\lambda_3 = 0$ during the first half, and $\lambda_3 = 10^{-2}$ in the second half. We initially set $\lambda_3 = 0$ due to the fourth-order derivative $u_{xxxx}$. We do not include a regularization term in this case $(\lambda_4 = 0)$ as the test loss was not significantly larger than the training loss. We train the networks for 3000 epochs, with an initial learning rate of $10^{-3}$ that decays exponentially. The network $\boldsymbol{\mathcal{E}}$ has dense layers of sizes [128, 128, 128, 32] with swish activation functions, followed by three linear layers of size $N_h$. The decoder $\boldsymbol{\mathcal{D}}$ has the same architecture in reverse order and without the linear layers.\\
\newline
\textbf{Kolmogorov Flow}: 
The 2D Navier-Stokes equations (NSE) for an incompressible fluid in non-dimensional form are given by
\begin{equation}
    \partial_t u + u\partial_x u + v\partial_y u = -\partial_x p + Re^{-1} \nabla ^ 2 u + f_x
\end{equation}
\begin{equation}
    \partial_t v + u\partial_x v + v\partial_y v = -\partial_y p + Re^{-1} \nabla ^ 2 v + f_y
\end{equation}
subject to the incompressibility constraint $\nabla \cdot \boldsymbol{u} = \boldsymbol{0}$, where $p$ is the pressure and $Re$ is the Reynolds number. This is the same formulation as in \cite{Chandler_Kerswell_2013, LucasYasuda2022, Parker_Schneider_2022}, with Kolmogorov forcing $(f_x, f_y) = (\sin4y, 0)$. The spatial domain is a doubly periodic box $(x,y)\in [0,2\pi)^2$. Taking the curl of the NSE yields the vorticity equation
\begin{equation}\label{eqn:vorticity_eqn}
    \partial_t \omega + u\partial_x \omega + v\partial_y \omega = Re^{-1} \nabla ^ 2 \omega + 4\cos 4 y
\end{equation}
where $\omega = (\nabla\times\boldsymbol{u)}\cdot\hat{\boldsymbol{z}}$. The normalized instantaneous dissipation $D$ and energy input rates $E$ are defined as
\begin{align}
    D &= (4\pi^2ReD_{lam})^{-1} \int_0^{2\pi}\int_0^{2\pi}\omega^2dxdy \\
    E &= (4\pi^2ReD_{lam})^{-1}\int_0^{2\pi}\int_0^{2\pi}uf_xdxdy
\end{align}
where $D_{lam} = Re/(2n^2)$ is the dissipation rate of the laminar solution. The continuous shift invariance $x \mapsto x + s,\; s\in\mathbb{R}$ results in an infinity of dynamically equivalent snapshots. Hence, we apply First Fourier Slicing \cite{Rowley2000, Marensi2023, Engel2023} on the data as a pre-processing step before training the autoencoder. This is done by working in the subspace where the $(k_x = 1, k_y = 0)$ Fourier mode is purely real: given a discretized vorticity snapshot $\boldsymbol{\omega}\in \mathbb{R}^{N_x\times N_y}$, with spatial FFT2 $\tilde{\boldsymbol{\omega}} = \mathcal{F}_{x,y}(\boldsymbol{\omega})\in \mathbb{C}^{N_x\times N_y}$, the sliced snapshot is given by $\boldsymbol{\omega}_s = \mathcal{F}_{x,y}^{-1} (\tilde{\boldsymbol{\omega}}e^{-i\boldsymbol{k}_x\phi})$, where $\phi = \mathrm{atan2}[\mathcal{I}(\tilde{\boldsymbol{\omega}}_{1,0}), \mathcal{R}(\tilde{\boldsymbol{\omega}}_{1,0})]$. The evolution of the phase $\phi$ can be obtained from the reconstruction equation \cite{Rowley2000}
\begin{equation}\label{eqn:rec_eqn}
    \partial_t\phi = \frac{\sin(\boldsymbol{x}) \cdot \partial_t\boldsymbol{\omega}_s}{\sin(\boldsymbol{x}) \cdot \partial_x\boldsymbol{\omega}_s}
\end{equation}
where $\boldsymbol{x}$ is the discretization of the $x$-interval $[0, 2\pi)$. The temporal evolution $\boldsymbol{F}_s$ within the slice is given by 
\begin{equation}\label{eqn:rhs_slice}
    \boldsymbol{F}_s(\boldsymbol{\omega}_s) = \partial_t\boldsymbol{\omega}_s - \frac{\sin(\boldsymbol{x}) \cdot \partial_t\boldsymbol{\omega}_s}{\sin(\boldsymbol{x}) \cdot \partial_x\boldsymbol{\omega}_s} \partial_x\boldsymbol{\omega}_s
\end{equation}
resulting in the latent dynamics to be
\begin{equation}
    \partial_t\boldsymbol{h} = \Big( \boldsymbol{U}^T_M + \textbf{J}_{\boldsymbol{\mathcal{E}}}(\check{\boldsymbol{\omega}}_s)\cdot\boldsymbol{U}^T\Big)\cdot\boldsymbol{F}_s(\check{\boldsymbol{\omega}}_s) 
\end{equation}
For more details on equations \ref{eqn:rec_eqn} and \ref{eqn:rhs_slice}, we refer to \citet{Rowley2000}. Convolutional layers can also serve as an alternative to slicing \cite{PBK2021, PageNorgaard2024}. As for the KSE, we train the network functions $\boldsymbol{\mathcal{E}}, \boldsymbol{\mathcal{D}}$ on the Fourier modes. Since the vorticity fields are real, multiple Fourier modes are complex conjugates of each other. Moreover, some entries are purely real, and applying a 2/3 de-aliasing results in 1847 unique degrees of freedom that serve as inputs to the autoencoder. We train the networks on a dataset of $\approx 10^6$ snapshots (10\% of which are reserved for the test dataset) for 3,000 epochs ($\approx1.5$ days on a H100 GPU) at float64 precision. In the first half we set $\lambda_{1} = 0.1,\ \lambda_{2} = 0.1,\ \lambda_{3} = 0\ \mathrm{and}\ \lambda_{4} = 10^{-9}$. In the second half we set $\lambda_{2} = 0.01,\ \lambda_{3} = 0.1$. We again keep $\lambda_{3} = 0$ in the first half to give the network time to learn the high-frequency terms, which dominate due to spatial derivatives. The network $\boldsymbol{\mathcal{E}}$ has dense layers of sizes [7000, 3000, 3000, 3000, 2000, 512] with swish activation functions, followed by three linear layers of size $N_h$. The decoder $\boldsymbol{\mathcal{D}}$ has the same architecture in reverse order and without the linear layers.

\subsection*{Computing invariant solutions} \label{sec:conv_inv_sol}
A fixed point $\boldsymbol{u}^*$ satisfies $\partial_t\boldsymbol{u}^* = 0$. We can find fixed points by searching for roots of the cost function
\begin{equation}
    \label{eqn:cost_fp}
    J_{fp}(\boldsymbol{u}) = ||\partial_t \boldsymbol{u}||
\end{equation}

To compute UPOs, we employ a variational loop convergence algorithm. An initial guess consists of a closed curve $\boldsymbol{u}(\boldsymbol{x},t)$ (a loop) in the state space and a guess period $T$, so that $\boldsymbol{u}(\boldsymbol{x},0) = \boldsymbol{u}(\boldsymbol{x},T)$. Hence, the loop is already time-periodic, but does not necessarily satisfy the flow equations. The loop is a UPO if and only if the flow vectors and tangent vectors align at every point of the loop. This misalignment is quantified by the cost function
\begin{equation}
    \label{eqn:cost_upo_integral}
    J_{upo}^2 = \int_0^1\bigg\rvert\bigg\rvert \frac{\partial \boldsymbol{u}}{\partial t} - \frac{1}{T}\frac{\partial \boldsymbol{u}}{\partial s}\bigg\lvert\bigg\lvert^2ds
\end{equation}
where $s$ is a parametrization of the loop, which in our case is the arclength parameter. A root of $J_{upo}$ then corresponds to a periodic orbit. Equations \ref{eqn:cost_fp} and \ref{eqn:cost_upo_integral} refer to the physical state $\boldsymbol{u}$. With a latent dynamics $\partial_t \boldsymbol{h}$, we can define analogous cost functions in the latent space. If we had perfect functions $\boldsymbol{\chi}$ and $\check{\boldsymbol{\chi}}$, then by construction of the latent dynamics from the chain rule (equation \ref{eqn:latent_dyn}), a fixed point in the full space corresponds to a fixed point in the latent space since $\partial_t\boldsymbol{u} = \boldsymbol{0} \implies \partial_t\boldsymbol{h} = \nabla_{\boldsymbol{u}}\boldsymbol{\chi}\cdot\partial_t\boldsymbol{u} = \boldsymbol{0}$. The converse is not necessarily true. Practically, given imperfect functions $\boldsymbol{\chi}$ and $\check{\boldsymbol{\chi}}$, we only know that 
\begin{equation}
    \partial_t\boldsymbol{h} = 0 \implies \partial_t\check{\boldsymbol{u}} \in \mathrm{Null}( \nabla_{\boldsymbol{u}} \boldsymbol{\chi}\rvert_{\boldsymbol{u} = \check{\boldsymbol{u}}})
\end{equation}

We make a similar observation for UPOs: a UPO of the full physical system satisfies 
\begin{equation}
\label{eqn:phys_upo}
    ||\partial_t\boldsymbol{u} - T^{-1}\partial_s\boldsymbol{u}|| = 0
\end{equation}
for all $t\in[0,T)$. In the latent space, this implies
\begin{equation}
    ||\partial_t\boldsymbol{h} - T^{-1}\partial_s\boldsymbol{h}|| = ||\nabla_{\boldsymbol{u}}\boldsymbol{\chi}\cdot(\partial_t\boldsymbol{u} - T^{-1}\partial_s\boldsymbol{u})|| = 0
\end{equation}

Therefore, a UPO in the physical space implies a UPO in the latent space. Again, the converse is not necessarily true. Since our neural networks are built with {\fontfamily{qcr}\selectfont Flax} \cite{flax2020github}, we also implement the computation of $J_{fp}$ and $J_{upo}$ in {\fontfamily{qcr}\selectfont JAX} \cite{jax2018github}. We can thus use automatic differentiation to find their roots. Note in particular that $\partial_t\boldsymbol{h}$ requires a derivative of a neural network, which is easily computed with automatic differentiation. To actually find roots of $J_{fp}$ and $J_{upo}$, we use {\fontfamily{qcr}\selectfont jaxopt}'s implementation of the quasi-Newton L-BFGS solver, and the Gauss-Newton solver \cite{jaxopt_implicit_diff} on the residual vector when necessary. Note that the activation function must be at least $C^2$ (e.g. the swish function), as the computation of the cost function in the latent space requires second derivatives of $\boldsymbol{h}$.

The initial guesses for the loop convergence algorithm are based on a latent recurrent flow analysis (RFA), which consists in generating a latent trajectory $\boldsymbol{h}(t)$, with $t\in[0, \ t_{max})$, and looking for sub-trajectories that almost close on themselves according to a near recurrence criterion $r$. Concretely, we look for $t_1, t_2$ such that the relative difference between $\boldsymbol{h}(t_1)$ and $\boldsymbol{h}(t_2)$ is $\epsilon(\boldsymbol{h}(t_1), \ \boldsymbol{h}(t_2)) < r$. The sub-trajectory $\boldsymbol{h}(t), \;t\in[t_1, t_2)$ then serves as a guess for the convergence algorithm, with guess period $T_0 = t_2 - t_1$. We smoothen the discontinuity by setting the temporal high-frequency terms to zero. 
The RFA for the KSE is conducted with recurrence criterion $r < 0.04$ and guess period $T_0\in[20,200]$. For Kolmogorov flow, the RFA for short recurrences were conducted with $T_0\in[2,50]$ and $r < 0.35$, while for the long ones we chose $T_0\in[50,80]$ with $r < 0.4$. The time-serieses we used were of either 3000 or 4000 time units. Including the discrete symmetries of the physical system (if present) in the recurrence function \cite{Chandler_Kerswell_2013} would optimize the approach and yield better guesses for pre-periodic orbits. However, in this paper, we do not implement this.

\section*{Acknowledgements}
The authors thank Jeremy P. Parker and Omid Ashtari for helpful discussions. This work was supported by the European Research Council (ERC) under the European Union’s Horizon 2020 research and innovation programme (grant no. 865677).

\bibliography{refs.bib}

\end{document}